\def\year{2019}\relax
\documentclass[letterpaper]{article} 
\usepackage{aaai19}  
\usepackage{times}  
\usepackage{helvet}  
\usepackage{courier}  
\usepackage{url}  
\usepackage{graphicx}  
\graphicspath{{./figs/}}
\usepackage{amsmath}
\usepackage{xcolor}

\def\PP{{{\rm l}\kern - .15em {\rm P} }}
\def\PN2{{\PP_{N}-\PP_{N-2}}}

\newcommand{\bphi}{\boldsymbol{\varphi}}

\newcommand{\btau}{\boldsymbol{\tau}}

\newcommand{\bbf}{\textbf{f}}

\newcommand{\ba}{\boldsymbol{a}}

\newcommand{\bb}{\boldsymbol{b}}

\newcommand{\bu}{\boldsymbol{u}}
\newcommand{\bU}{\boldsymbol{U}}

\newcommand{\bx}{\boldsymbol{x}}
\newcommand{\bX}{\boldsymbol{X}}

\newcommand{\deleted}[1]{{}}

\usepackage{algorithm,algpseudocode}
\usepackage{booktabs,multirow}
\usepackage{diagbox}
\usepackage{bigints}

\begin{document}

\title{%
Non-intrusive inference reduced order model for fluids using linear multistep neural network
}

 \author{Xuping Xie, Guannan Zhang, Clayton G.~Webster\\
 \{xiex, zhangg, webstercg\}@ornl.gov\\
 Computer Science and Mathematics Division, Oak Ridge National Laboratory\\
 One Bethel Valley Road, Oak Ridge, TN 37831
 }

\maketitle
\begin{abstract}
In this effort we propose a data-driven learning framework for reduced order modeling of fluid dynamics. Designing accurate and efficient reduced order models for nonlinear fluid dynamic problems is challenging for many practical engineering applications.  Classical projection-based model reduction methods generate reduced systems by projecting full-order differential operators into low-dimensional subspaces.  However, these techniques usually lead to severe instabilities in the presence of highly nonlinear dynamics, which dramatically deteriorates the accuracy of the reduced-order models. In contrast, our new framework exploits linear multistep networks, based on implicit Adams-Moulton schemes, to construct the reduced system. The advantage is that the method optimally approximates the full order model in the low-dimensional space with a given supervised learning task. Moreover, our approach is non-intrusive, such that it can be applied to other complex nonlinear dynamical systems with sophisticated legacy codes. We demonstrate the performance of our method through the numerical simulation of a two-dimensional flow past a circular cylinder with Reynolds number Re = 100. The results reveal that the new data-driven model is significantly more accurate than standard projection-based approaches.

\end{abstract}

\section{Introduction}
\label{sec:intro}
The full order model (FOM) of realistic engineering applications in fluid dynamics often represents a large scale dynamic system. High-fidelity CFD (computational fluid dynamics) simulations of the FOM are so computationally expensive that they put a heavy burden on the computational resources despite the available CFD software and supercomputers with thousands of cores. Consequently, the use of FOM for such simulations is often impractical and prohibitive for time-critical applications such as system identification, flow control, design optimization.

 The reduced order modeling in fluid dynamics is to construct an accurate low-dimensional approximation to the full system with orders of magnitude reduction in computational cost. The first usage of reduced order model (ROM) in fluid dynamics was by \cite{lumley1967structure} for studying the intensity of turbulence and coherent structures. Many recent successful applications of ROM in fluid problems can be found in \cite{noack2011reduced,obinata2012model,carlberg2013gnat,rowley2017model,amsallem2012stabilization,xie2017approximate,kutz2016dynamic,san2018extreme}

The Galerkin projection based reduced order model (GP-ROM) is one of the most popular methods that has been widely used in practice. The GP-ROM, in an offline stage, first constructs a reduced space and then uses the Galerkin projection of the FOM operator to obtain a low-dimensional nonlinear dynamical system, i.e., ROM dynamics. The reduced space is often generated by proper orthogonal decomposition (POD), also known as principal component analysis.  In an online stage, the obtained reduced dynamics can be used to approximate the full system efficiently for various applications, such as long-term prediction, flow control. However, the projection step requires that the full model operators have to be available in order to obtain the ROM dynamics. This limits the applicability of projection based model reduction in situations where the full model is unknown \cite{xiao2015non,peherstorfer2016data}. More importantly, the computational cost of assembling the reduced operators -- tensors from the projection of FOM operator -- scales with the large dimension of the underlying high-dimensional of FOM. For this reason, the GP-ROM are efficient for problems where the reduced operators must be constructed only once.

On the other hand, the GP-ROM generates inaccurate approximations for highly non-stationary (nonlinear) fluids, e.g., turbulence. In the literature, a common explanation for this failure is that the Galerkin projection does not preserve the stability properties from the full model. A deeper reason is that the low-dimensional space used in the Galerkin projection cannot resolve the nonlinear interaction of the fluid system \cite{noack2016recursive,loiseau2018constrained}. Consequently, resulting in a projection based stability error which makes GP-ROM fail in nonlinear fluid applications, e.g., see \cite{balajewicz2013low,ballarin2015supremizer,carlberg2015galerkin,xie2018data}

\subsection{Related Work}
\textbf{Closure modeling.} Numerous stabilization strategies have been devised to address the instability problem, known as closure modeling. The fundamental idea of closure modeling is to model the lost information from the low-dimensional space since it is generated through POD truncation. This truncation is to keep the first few POD modes that extract the most dominant structure of the full system and discard the rest modes. Closure models generally can be categorized into two common approaches. 
One is physically model the effect of the discarded POD mode by adding artificial viscosity to the reduced system~\cite{balajewicz2013low,ballarin2015supremizer,protas2015optimal,amsallem2012stabilization}. Another approach is to mathematically model the ROM dynamics by solving a related optimization problem \cite{carlberg2015galerkin,xie2018data}. The most recent development for closure modeling is to apply the neural network to approximate the lost information \cite{san2018neural}.


\textbf{Differential equations learning.} Bridging numerical differential equations and deep neural networks have gained enormous attention recently. Specifically, \citeauthor{chang2017multi} (2017) proposed dynamical system viewing of residual networks (ResNet). \citeauthor{lu2017beyond} (2017) the first time introduced the linear multistep network architecture to analyze ResNet on classification task. Sparse regression, Gaussian process, multistep neural networks have been applied for dynamic system learning \cite{brunton2016discovering,rudy2017data,raissi2018hidden,raissi2017machine}. 
More recently, ordinary differential equation network (ODE-net) was introduced for supervised learning \cite{chen2018neural}.

\subsection{Our approach}
In this paper, we propose a novel non-intrusive learning reduced order model framework for fluid dynamic system. The new framework provides a general concept of learning the optimal reduced dynamic system from the data. Inspired by the successful development in learning differential equations with deep networks, we apply the linear multistep neural network (LMNet) to learn the reduced order model (LMNet-ROM). Unlike closure modeling and the existed non-intrusive model reduction methods, in this work, we focus on a different perspective. First, we do not use Galerkin projection and model the closure problem. The optimal reduced dynamic system that can address the instability issue is learned for a given supervised learning task. Secondly, the new model does not approximate reduced operators whereas other non-intrusive models use interpolation or regression method to infer reduced operators \cite{peherstorfer2016data,xiao2015non}. Moreover, our viewpoint easily enables us to answer the common question -- what is the best ROM dynamic system to approximate the full system for a given low-dimensional subspace? We demonstrate the performance of the new LMNet-ROM is better than GP-ROM in full order model approximation and long-term prediction. The main contributions can be summarized as follows:
\begin{itemize}
\item A novel non-intrusive learning reduced order model framework for fluid dynamics, which is applicable to general nonlinear dynamical systems with sophisticated legacy codes.
\item Our framework overcomes the instability issue of
the projection based model reduction, and provides accurate approximation and long-term prediction of the full system.
\item The learning process of our approach is more computationally efficient than the construction of reduced operators in the classic projection based methods.
\end{itemize}

\section{Reduced Order Modeling}
\label{ROM}
In this section, we present the Galerkin projection based reduced order modeling framework in fluid dynamic system. The classical Navier-Stokes equations (NSE) are often used as a mathematical model in fluid dynamics,
\begin{eqnarray}
    && \frac{\partial \bu}{\partial t}
    - Re^{-1} \Delta \bu
    + \bu \cdot \nabla \bu
    + \nabla p
    = {\bf 0} \, ,
    \label{eqn:nse-1}                                                         \\
    && \nabla \cdot \bu
    = 0 \, ,
    \label{eqn:nse-2}
\end{eqnarray}
where $\bu$ is the velocity, $p$ the pressure, and $Re$ the Reynolds number. 
We use the initial condition $\bu(\bx, 0) = \bu_0(\bx)$ and (for simplicity) homogeneous Dirichlet boundary conditions: $\bu(\bx, t) = {\bf 0}$. For convenience, we use $\frac{\partial\bu}{\partial t} = \bbf(\bu, Re)$ as the general notation for the NSE in the rest of the paper. 

\textbf{Reduced Space.} Proper orthogonal decomposition (POD) is the dimension reduction method that we used to generate the reduced space. It starts with the data matrix, $\bU=[\bu^0,\bu^1,\cdots,\bu^s]\in\mathcal{R}^{D_h\times (s+1)}$, collected by the numerical solutions or experimental observations of the full system \eqref{eqn:nse-1} at $s+1$ different time instances. The POD method seeks a low-dimensional space $\bX^r$ that approximates the data ($\bU$) optimally with respect to the $L^2$-norm. It formulates the following eigenvalue problem:
\begin{eqnarray}
\label{eqn:pod-svd}
\bU\bU^T\varphi_i = \lambda_i \varphi_i, \,\,\, i=1,2,...,d
\end{eqnarray}
where $d$ is the rank of the data matrix $\bU\bU^T$. $\lambda_i$ and $\varphi_i$ are the eigenvalues and POD basis, respectively.
The reduced space is given after the truncation as $\Phi_r:=\{\varphi_1,\cdots,\varphi_r\}\in\mathcal{R}^{D_h\times r}$. 

\textbf{Galerkin Projection-ROM (GP-ROM).} For a given space $\Phi_r$, the GP-ROM finds the approximation of the velocity field spanned by the low-dimensional space, 
\begin{equation}
	\bu\approx {\bu}_r({\bf x},t) 
	\equiv \sum_{j=1}^r a_j(t) \bphi_j({\bf x}) \, ,
	\label{eqn:g-rom-1}
\end{equation}
 where $\{a_{j}(t)\}_{j=1}^{r}$ are the sought time-varying coefficients. The GP-ROM can be obtained by projecting the FOM onto the POD space: $\forall \, i = 1, \ldots, r,$
    \begin{eqnarray}
        \left(
            \frac{\partial \bu_r}{\partial t} , \bphi_{i}
        \right)
        = \left( \bbf(\bu_r, Re), \bphi_i\right).
    \label{eqn:g-rom-weak}
    \end{eqnarray}
Here, $(\cdot,\cdot)$ is the $L^2$ inner product. The solution of GP-ROM can be determined by the following nonlinear dynamic system:
\begin{eqnarray}
	\dot{\ba}
	= L \, \ba
	+ \ba^{\top} \, N \, \ba \, ,
	\label{eqn:g-rom}
\end{eqnarray}
where $L$ and $N$ are ROM operators that can be obtained by projection
\begin{eqnarray}
	 L_{im}
	=  \frac{1}{Re} \, \left( \Delta \bphi_m , \bphi_i \right),
	 N_{imn}
	= - \bigl( \bphi_m \cdot \nabla \bphi_n , \bphi_i \bigr)
	\label{eqn:g-rom-3b}
\end{eqnarray}
Note that the reduced system \eqref{eqn:g-rom} efficiently approximates the full model of NSE as the dimension $r$ is generally very small ($\sim\mathcal{O}(10)$) compared to the high dimension of $\bu$ ($D_h\sim\mathcal{O}(\geq 10^5)$).

\textbf{ROM closure models.} 
The ROM closure models can be generally written as the following dynamic system: 
\begin{eqnarray}
\dot{\ba}
	= L \, \ba
	+ \ba^{\top} \, N \, \ba \, + \btau  ,
	\label{eqn:rom-closure}
\end{eqnarray}
$\btau$ is an artificial term that model the effect of the discarded POD modes using various approaches e.g., see \cite{parish2016paradigm,gouasmi2017priori,wells2017evolve,xie2018data,san2018neural}. 
We note that the dynamics of closure model \eqref{eqn:rom-closure} is more accurate than GP-ROM dynamics \eqref{eqn:g-rom} in approximation of the model. The challenge of closure modeling is that the $\btau$ is unknown, i.e., no explicit formula. Therefore, closure models are empirical modification of the GP-ROM from physical or mathematical perspective.

\section{Learning Reduced Order Model}
\label{sec:MNN}
In this section, we present the architecture of learning the reduced order model from deep neural networks. In contrast to the standard GP-ROM framework, we learn the reduced dynamical system from the data without intrusively using ROM operators (e.g., $L,N$)

\subsection{Optimal ROM dynamics}
\label{sec:learning-rom}
We consider the low dimensional ROM dynamic system as a general function,
\begin{eqnarray}
\label{eqn:rom-dy}
\dot{\ba} = \bbf_r(\ba).
\end{eqnarray}
We claim that this function $\bbf_r$ has a general representation of the ROM dynamics including system \eqref{eqn:g-rom} and \eqref{eqn:rom-closure}. Our goal is to learn the ROM dynamics \eqref{eqn:rom-dy} in a given set of temporal data and return a closed form model that can be used to accurately approximate and predict the full system. 

For a given data-set of snapshot solutions of NSE \eqref{eqn:nse-1}, $\bU=\bu^0,...,\bu^s\in \mathcal{R}^{D_h\times{s+1}}$, at time steps $t_0,...,t_s$. The best approximation of snapshots data by the POD space is given by, $\bu^j = \sum_{i=1}^rb_i^j\varphi_i$, $j=t_0,...,t_s$. The reduced dynamics \eqref{eqn:rom-dy} is to find coefficients $\ba$ such that,
\begin{eqnarray}
\bu_r = \sum_{i=1}^ra_i^j\varphi_i \approx \bu = \sum_{i=1}^rb_i^j\varphi_i
\end{eqnarray}
It indicates that the optimal solution from the $r$-dimensional ROM dynamic system is given by the full model data, $\bb^j$, such that
\begin{eqnarray}
\label{eqn:dns-rom}
\dot{\bb} = \bbf_r(\bb)
\end{eqnarray}
This provides a framework that learning the ROM dynamics from the data-set $B=[\bb^0,...,\bb^s]\in \mathcal{R}^ {r\times (s+1)}$, i.e., time-varying coefficients of FOM data. The training data-set can be computed by the following,
\begin{eqnarray}
\label{eqn:training-mat}
B = \Phi_r^TW\bU
\end{eqnarray}
The formula is derived by the $L^2$ projection of data $\bU$ to the low-dimensional space $\Phi_r$. $W\in\mathcal{R}^{D_h\times D_h}$ is the weight matrix of the $L^2$ inner product, we use finite element weights in this work.

\subsection{Linear Multistep Network (LMNet)} 
Motivated by differential equation learning, we adopt the linear multistep network architecture \cite{lu2017beyond,raissi2018multistep} to construct a structured nonlinear regression model that can learn the reduced dynamics. In this work, we only consider the implicit multistep method, Adams-Moulton (AM) scheme \cite{ascher1998computer} as it has better stability property. The $K$-step AM method is defined as follows:
\begin{eqnarray}
\label{eqn:ode-am}
\ba^n = \sum_{i=1}^K(\alpha_i\ba^{n-i}+\beta_i\Delta t\bbf_r(\ba^{n-i}))+\beta_0\Delta t \bbf_r(\ba^n)
\end{eqnarray}
We discretize the ROM dynamic system \eqref{eqn:rom-dy} by using AM scheme \eqref{eqn:ode-am} with a neural network. The parameters of this neural network can be learned by minimizing the mean squared error loss function:
\begin{eqnarray}
\label{eqn:nn-lost}
MSE: = \frac{1}{N - K + 1}\sum_{n=K}^N|\mathcal{L}_n|^2
\end{eqnarray}
Where $S$ is the total number of time instance in the ROM dynamic system. $\mathcal{L}_n$ is the local truncation error from the Taylor expansion of $K$-step AM method \eqref{eqn:ode-am},
\begin{eqnarray}
\label{eqn:nn-lost-y}
\mathcal{L}_n = \sum_{i=0}^K\alpha_i\ba^{n-i}+\Delta t \beta_i\bbf_r(\ba^{n-1}).
\end{eqnarray}
One advantage of this nonlinear regression is that we do not have to approximate the temporal gradients \cite{peherstorfer2016data,rudy2017data} since the time derivatives are discretized by the AM method.

\subsection{LMNet-ROM}
\begin{figure}[h!]
              \includegraphics[width=\linewidth]{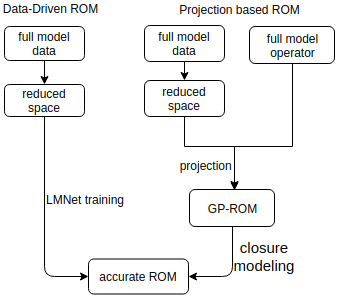}
              \caption{Flowchart of projection based model reduction and the new non-intrusive learning reduced order modeling framework}
\label{fig-nnirom-pic}
\end{figure}

We use the trained neural network as the ROM dynamic system \eqref{eqn:rom-dy} to approximate the full system of NSE. We emphasize that the novelty of our approach is the non-intrusively learning of the reduced system, whereas GP-ROM and closure models require the use of FOM operators. Fig.\ref{fig-nnirom-pic} is the flowchart of the LMNet-ROM and GP-ROM framework.
We outline the algorithm of the framework in the following:
\begin{algorithm}[H]
	\caption{Linear Multistep Network Reduced Order Model Learning (LMNet-ROM)}
	\label{alg:nni-rom}
	\begin{algorithmic}[1]
		\State{Compute the reduced POD space from the data of NSE by \eqref{eqn:pod-svd}	
		}
		\State{ Compute the training dataset $B$ via \eqref{eqn:training-mat}
					}
		\State{
				Train the neural network using loss function \eqref{eqn:nn-lost}
					}		
		 \State{
					The LMNet-ROM for NSE is obatined from the trained low-dimensional dynamic system:
								\begin{equation}
									\boxed{
									\dot{\ba}={\bbf}_r^{Net}(\ba)
									} 
									\label{eqn:mnni-rom-online}
								\end{equation}
					}
\end{algorithmic}
\end{algorithm}
\noindent We claim that the learned reduced dynamics \eqref{eqn:mnni-rom-online} has a better approximation to the full model than system \eqref{eqn:g-rom-3b} and \eqref{eqn:rom-closure} since it is learned optimally from the FOM data. The new framework, see Fig.\ref{fig-nnirom-pic}, only requires input data from a system and does not use any FOM operator, which can be generally applied to reduced order modeling of any fluid dynamical system. The main offline computational cost of the LMNet-ROM is training the neural network whereas the GP-ROM is the construction of ROM operators in \eqref{eqn:g-rom-3b}. In the numerical experiment, we show that the offline computation of our model is faster than the GP-ROM. 

\section{Numerical Experiment}

In this section, we present the preliminary numerical results to demonstrate the advantages of our new model. The test case is a 2D channel flow past a circular cylinder at a $Re=100$. It is a benchmark problem that has been used as a numerical test in fluid dynamics, see \cite{schaefer1996benchmark,kutz2016dynamic,brunton2014compressive,mohebujjaman2017energy,xie2018data}.

\subsection{Implementation Details}
	\label{sec:test-problem-setup}

\begin{figure}[htp]
\begin{center}
\includegraphics[width=0.49\textwidth,height=0.2\textwidth, trim=0 0 0 0, clip]{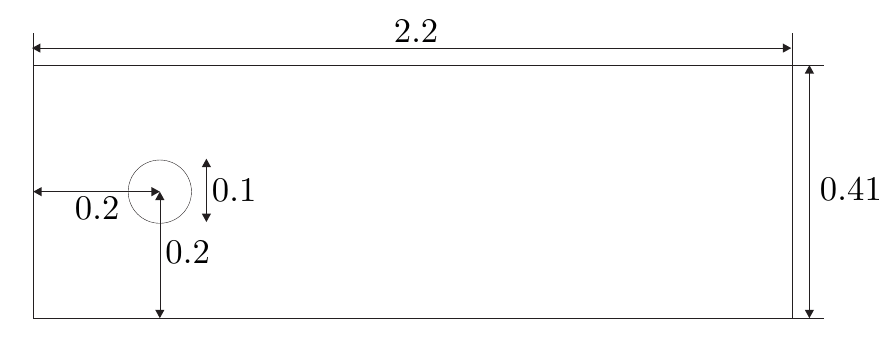}
\end{center}
\caption{\label{cyldomain} Channel flow around a cylinder domain.}
\end{figure}

The domain is a $2.2\times 0.41$ rectangular channel with a radius=$0.05$ cylinder, centered at $(0.2,0.2)$, see Figure~\ref{cyldomain}.  
No slip boundary conditions are prescribed for the walls and on the cylinder, and the inflow and outflow profiles are given by~\cite{mohebujjaman2017energy,xie2018data} $u_{1}(0,y,t)=u_{1}(2.2,y,t)=\frac{6}{0.41^{2}}y(0.41-y) \, , u_{2}(0,y,t)=u_{2}(2.2,y,t)=0$.
The kinematic viscosity is $\nu=10^{-3}$, there is no forcing, and the flow starts from rest.

The velocity snapshots of NSE \eqref{eqn:nse-1} are generated by finite element method with meshsize approximately 103K which gives a fully resolved solution. The lift and drag computation agree well with results from references in~\cite{caiazzo2014numerical,ST96}: $c_{d,max} = 3.2261,\ c_{l,max} = 1.0040.$. A total number of 2500 snapshots $\bU$ are collected from $T=[0, 5]$ at every time step $\Delta t= 0.002$. The LMNet-ROM is built and trained in Tensorflow.

\subsection{Full Order Model Approximation}
After obtaining the trained neural network, we use the ROM dynamic system \eqref{eqn:mnni-rom-online} to approximate the full model of NSE \eqref{eqn:nse-1}. The following average $L^2$ error formula has been used to quantify the accuracy of the model, 
\begin{eqnarray}
\label{eqn:err}
\mathcal{E}=\frac{1}{s+1}\sqrt{\sum_{j=0}^s\bigintsss_\Omega (\bu^j-\bu^j_r)d\Omega}.
\end{eqnarray}

We first evaluate the model with different layers and neurons. The dimension of the ROM dynamics is fixed to be $r=8$ with the one step Adams Moulton method.  Table \ref{table:layers} provides a crude estimate that increasing network width (256 neurons) might have the potential over-fitting issue whereas decreasing units (64) may not be enough to reach a good accuracy. Also, the network depth (number of hidden layers) have a positive effect on the performance of the model. To reduce the numerical efforts, we use one hidden layer and 128 neurons for the neural network in the rest of our evaluations. We emphasize that to fully understand the model sensitivity with respect to network architecture, a systematic study involving regularization, batch normalization and drop out is needed in future research. 
\bgroup
\def\arraystretch{1.3}
\begin{table}[h]
\caption{Average $L^2$ error between trajectories of the learned reduced order model with dimension $r=8$ and the exact data for the different number of hidden layers and neurons}
\vspace*{2mm}
\begin{tabular}{c|ccc}
\hline
\backslashbox{layers}{neurons}& 64 & 128 & 256 \\ \hline 
           1            & 5.62e-02 & 7.22e-03 & 4.31e-01 \\ \hline
           2           & 4.01e-03 & 5.75e-03 & 2.95e-02 \\ \hline
           3           & 1.64e-02 & 4.04e-03 & 3.11e-03 \\ \hline
\end{tabular}
\label{table:layers}
\end{table}
\egroup

We also test the LMNet-ROM with different steps in the training.
Table \ref{table:lm_method} lists the error between the new model and the exact data. As a comparison, we add GP-ROM result in the last entry. Table \ref{table:lm_method} shows that the LMNet-ROM consistently provide more accurate results as increasing steps of Adams Moulton (AM) method. An intuitive explanation is that the stability property of AM method helps to regularize the network and eventually achieve a good calibration. Large steps ($K$) requires high computational cost of training the network. To balance the output accuracy and training cost, we use step $K=1$ for the rest of our numerical tests. 

\bgroup
\def\arraystretch{1.3}
\begin{table}[h]
\caption{Average $L^2$ error between the new model and exact data for the different number of steps and dimensions}
\vspace*{2mm}
\begin{tabular}{c|ccc}
\hline
\backslashbox{K/model}{dimension} & r=4 & r=6 & r=8  \\ \hline
         1             &1.88e-03 & 3.05e-03e & 7.22e-03 \\ \hline
         2             &2.83e-04 & 6.04e-04 & 1.45e-03 \\ \hline
         3             &2.67e-04 & 6.22e-04 & 7.14e-03 \\ \hline
         4             & 3.79e-04 & 6.20e-04 & 6.37e-04 \\ \hline
         GP-ROM             & 1.66e-01 & 7.30e-02 & 1.92e-02 \\ \hline
\end{tabular}
\label{table:lm_method}
\end{table}
\egroup

 \bgroup
\def\arraystretch{1.3}
\begin{table}[h!]
\centering
\caption{Average $L^2$ error for different noise magnitudes. }
\vspace*{2mm}
\begin{tabular}{c|cc}
\hline
\backslashbox{noise}{model} & GP-ROM & LMNet-ROM\\ \hline
         0.0\%             & 1.92e-02 & 7.22e-03 \\ \hline
         0.5\%  & 1.93e-02 & 2.44e-03 \\ \hline
         1\%  & 2.05e-02 & 2.20e-02 \\ \hline
         5\%  & 2.92e-02 & 8.76e+01 \\ \hline
\end{tabular}
\label{table:noise}
\end{table}
\egroup

\textbf{Noisy data.} 
The above numerical tests are carried out on the deterministic data. In some situations, however, problems may contain noise measurements. We study the robustness of the new method with respect to noise data by adding Gaussian noise to the data-set for both models. Table \ref{table:noise} lists the error comparison between the LMNet-ROM and GP-ROM for different level of noise data. The results show that the LMNet-ROM cannot preserve good performance when the noise magnitude ($\geq$1\%) is high while GP-ROM does. The argument is that the learned dynamical system fully depends on the data which makes it vulnerable to noise interference. The GP-ROM, however, requires the use of FOM operators making it less sensitive than LMNet-ROM to noise. Further approaches should be investigated for the potential improvement for this problem. As for deterministic data, the LMNet-ROM is better than GP-ROM in full system approximation.

\subsection{Long-Term Prediction }
\begin{figure}[h!]
\centering
\includegraphics[width=0.4\textwidth]{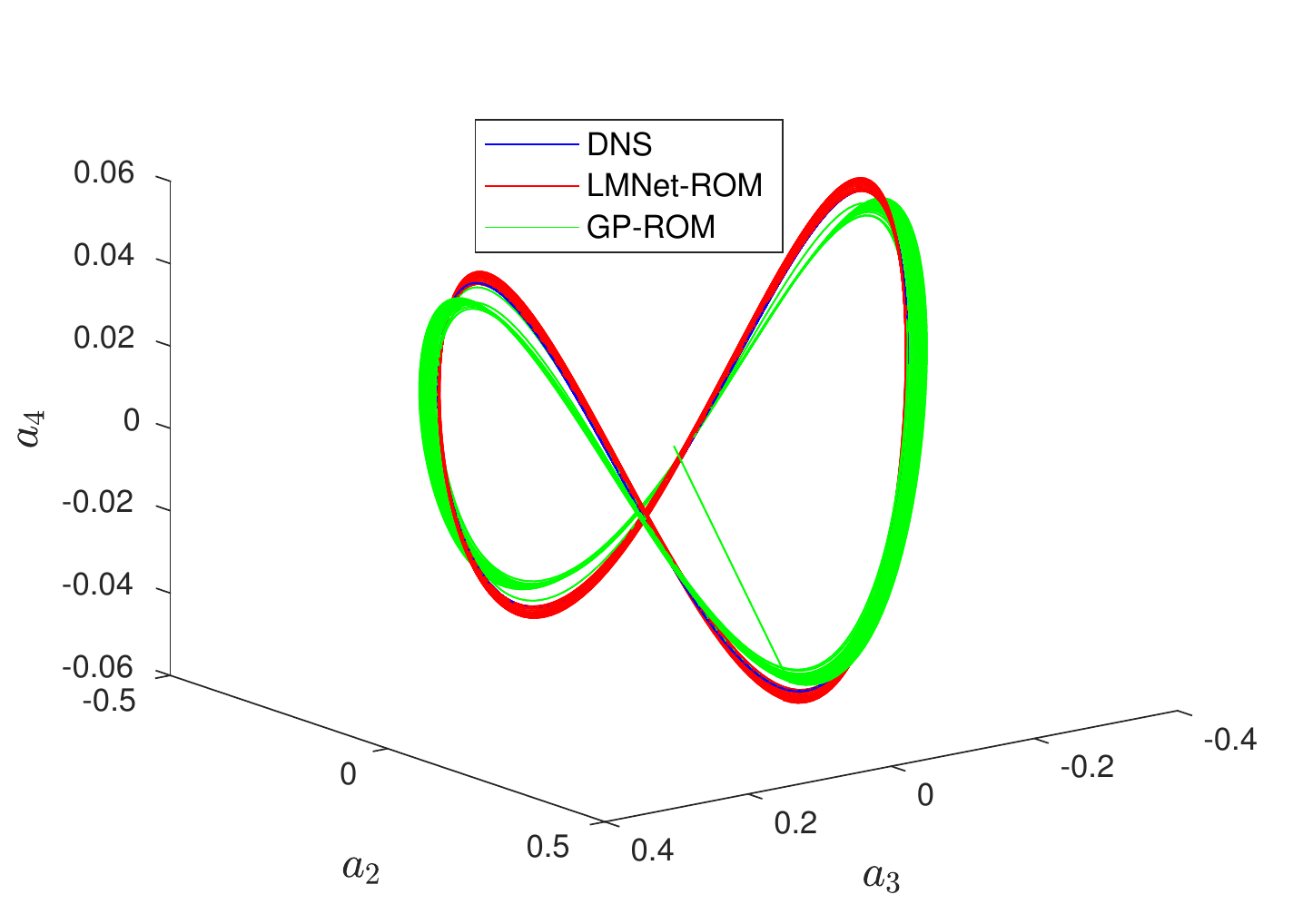}
\caption{phase portraits of the coefficients $a_2,a_3,a_4$ from LMNet-ROM (red), GP-ROM (green) and DNS data (blue) with dimension $r=8$}
\label{fig:phase}
\end{figure}

In this section, we make a thorough study of long-term predictability of the new LMNet-ROM. The solution to our new model and the GP-ROM are computed by the dynamic system \eqref{eqn:mnni-rom-online} and \eqref{eqn:g-rom}, respectively. We use direct numerical simulation (DNS) to denote the exact solution (data) of NSE. We use snapshots data that are collected from the time interval $T=[0,3]$ to generate the POD space and train the neural network. We then run the reduced systems \eqref{eqn:mnni-rom-online} and \eqref{eqn:g-rom} for $T=[0,5]$ to make the prediction. Fig. \ref{fig:phase} plots the phase portraits of the first few coefficients, $a_2, a_3, a_4$, from both models and the DNS data. The red line that depicts the result from LMNet-ROM has a closer mimic of DNS data, whereas the portraits from the GP-ROM have small deviation. This behavior tells that the dynamics of LMNet-ROM predict future states better than GP-ROM with the given information. Note that $a_1$ is a constant and not meaningful to discuss since the first POD mode $\varphi_1$ represents the mean flow. 

\begin{figure}[h!]
\centering
\begin{minipage}[c]{0.4\textwidth}
\includegraphics[width=\textwidth]{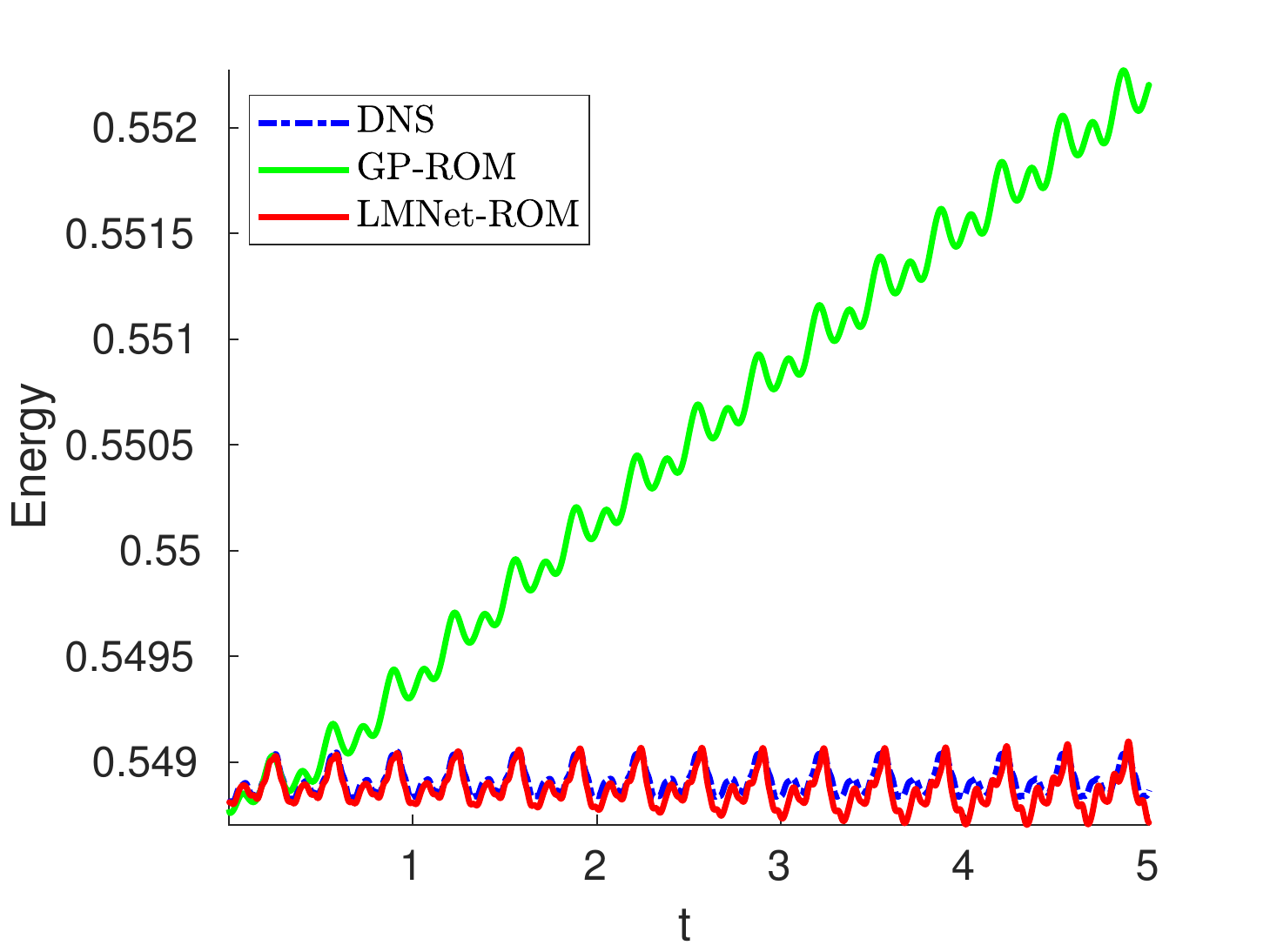}
\end{minipage}
\begin{minipage}[c]{0.4\textwidth}
\includegraphics[width=\textwidth]{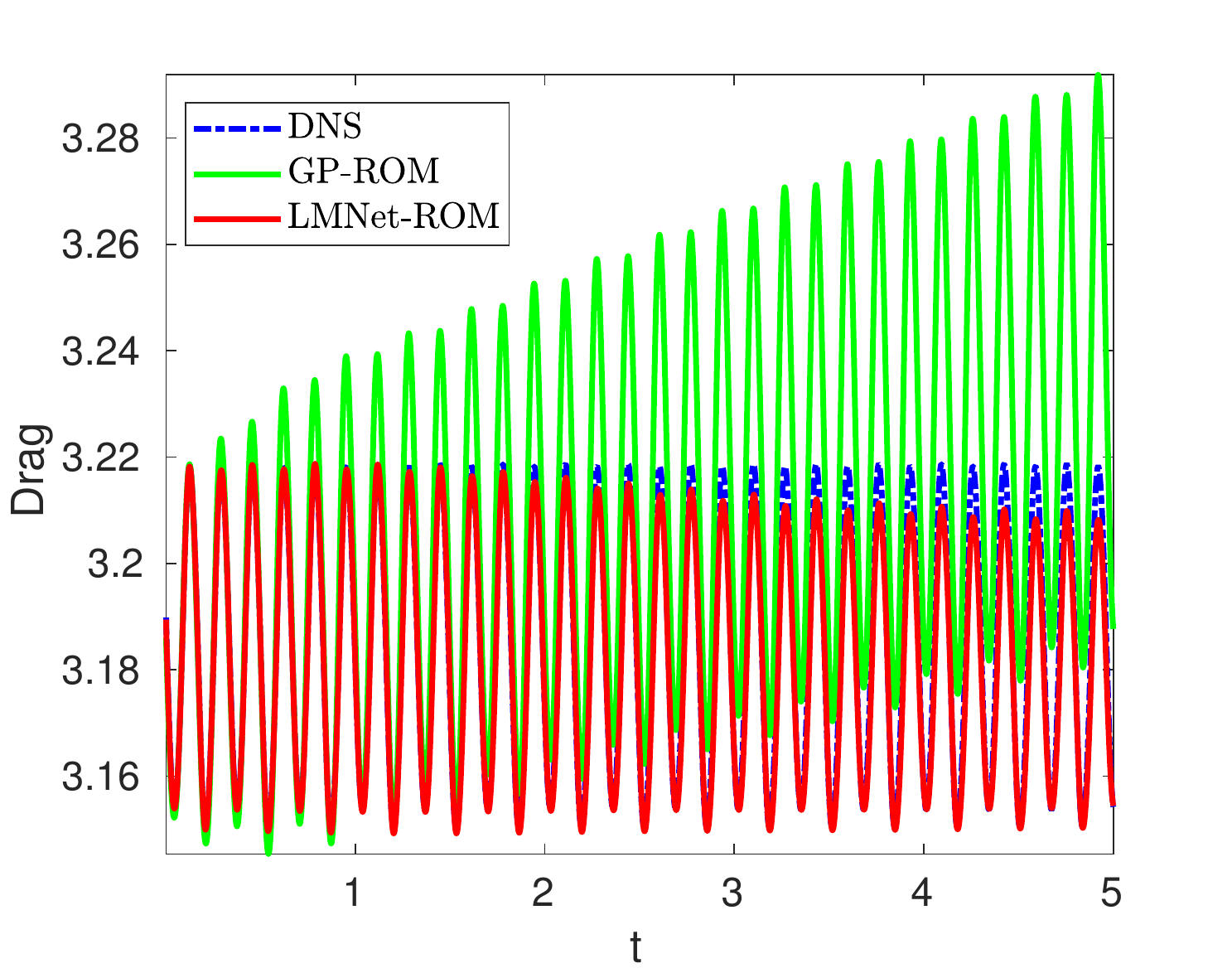}
\end{minipage}
\caption{Plots of the time evolution of energy $E(t_j)$ (top) and drag (bottom). The  solutions are generated from LMNet-ROM and GP-ROM with dimension $r=8$}
\label{fig:ener}
\end{figure}

\begin{figure*}[h!]
\includegraphics[width=0.33\textwidth]{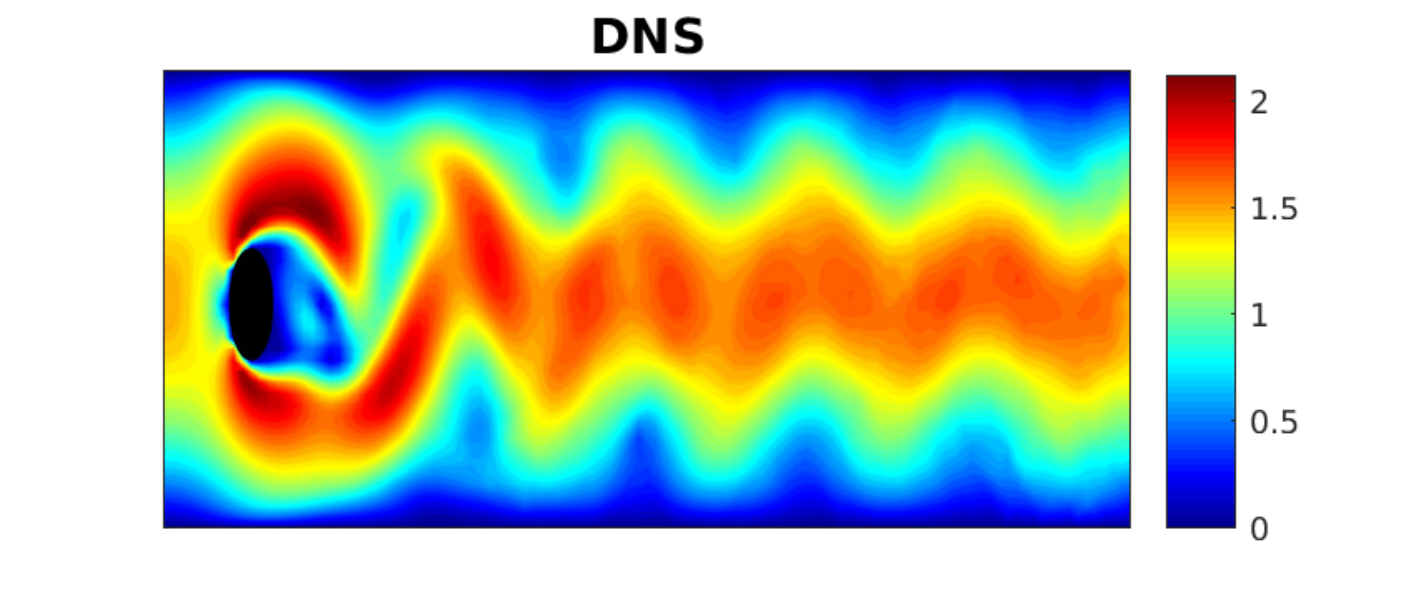}
\includegraphics[width=0.33\textwidth]{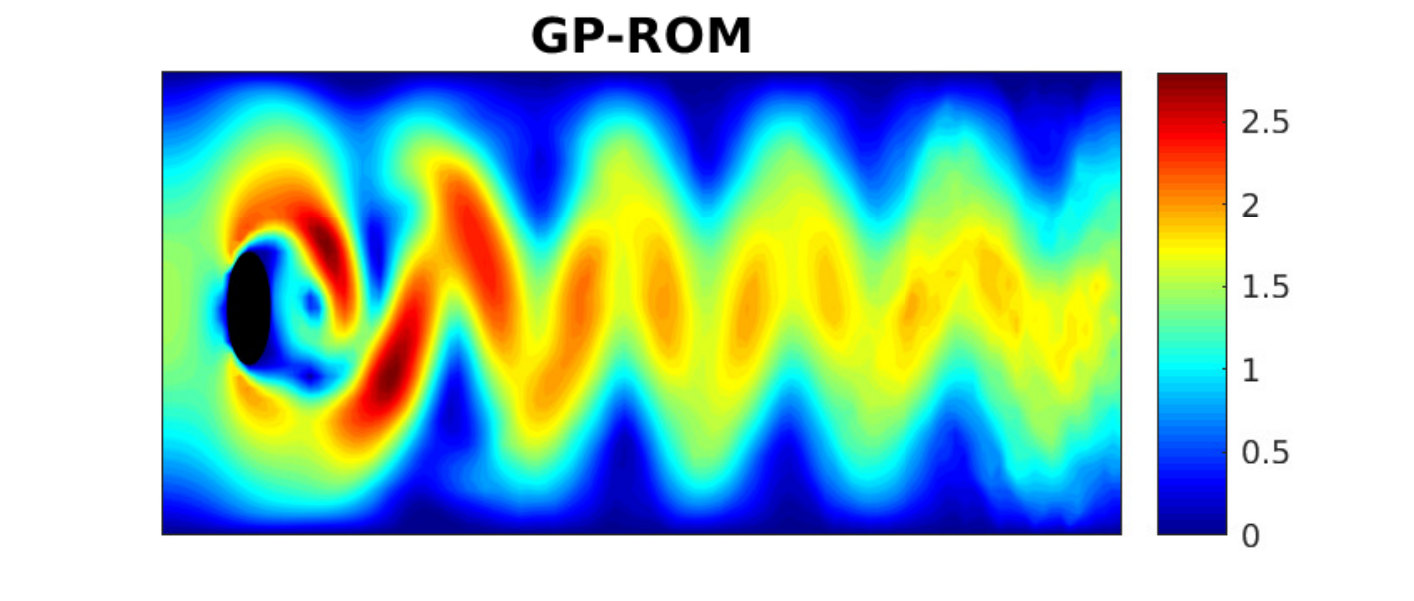}
\includegraphics[width=0.33\textwidth]{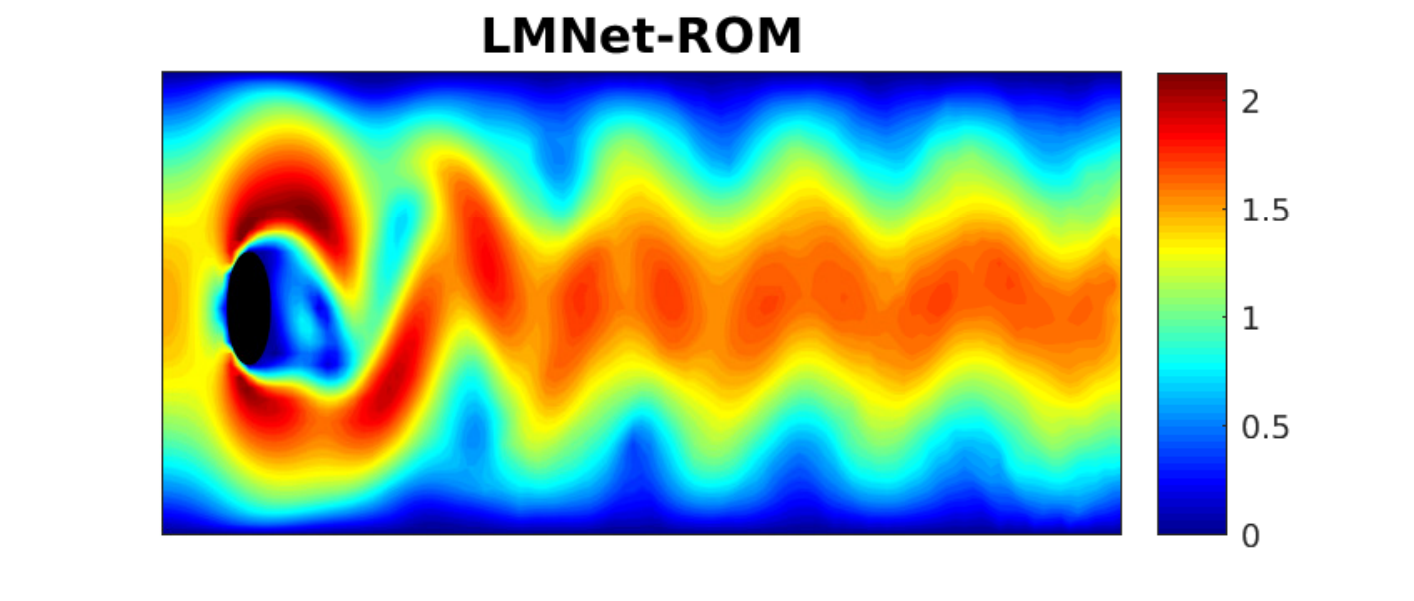}\\
\hspace*{-2mm}
\includegraphics[width=0.33\textwidth]{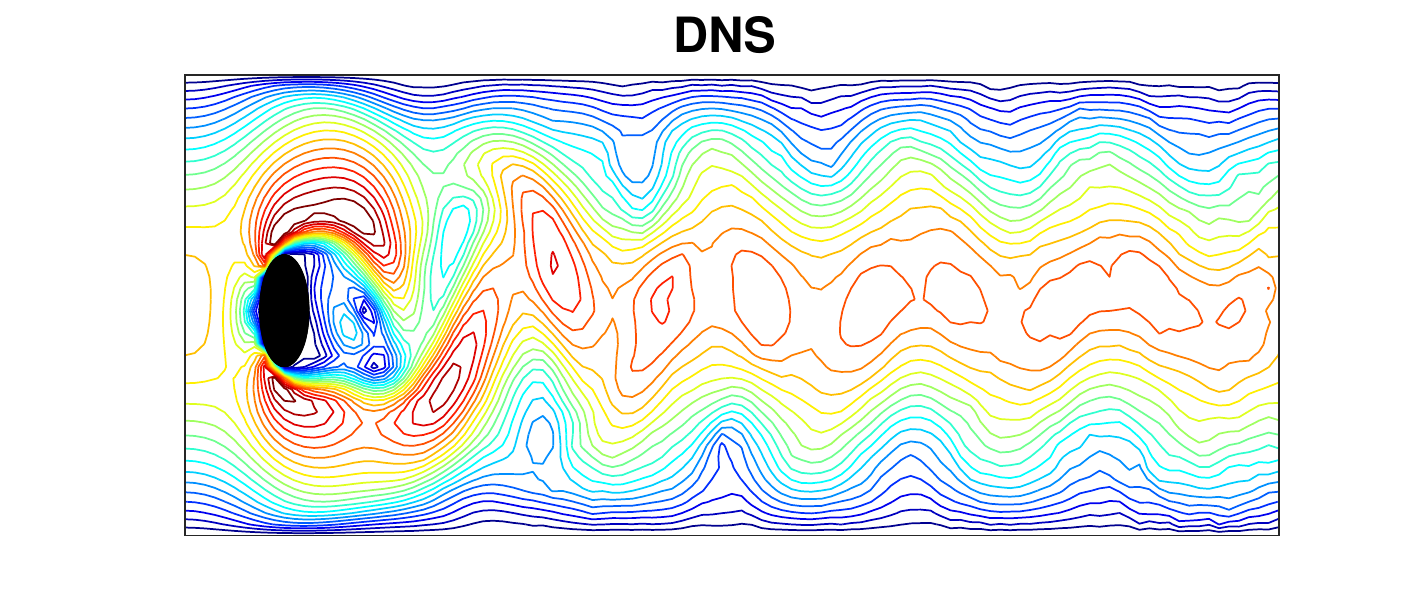}
\includegraphics[width=0.33\textwidth]{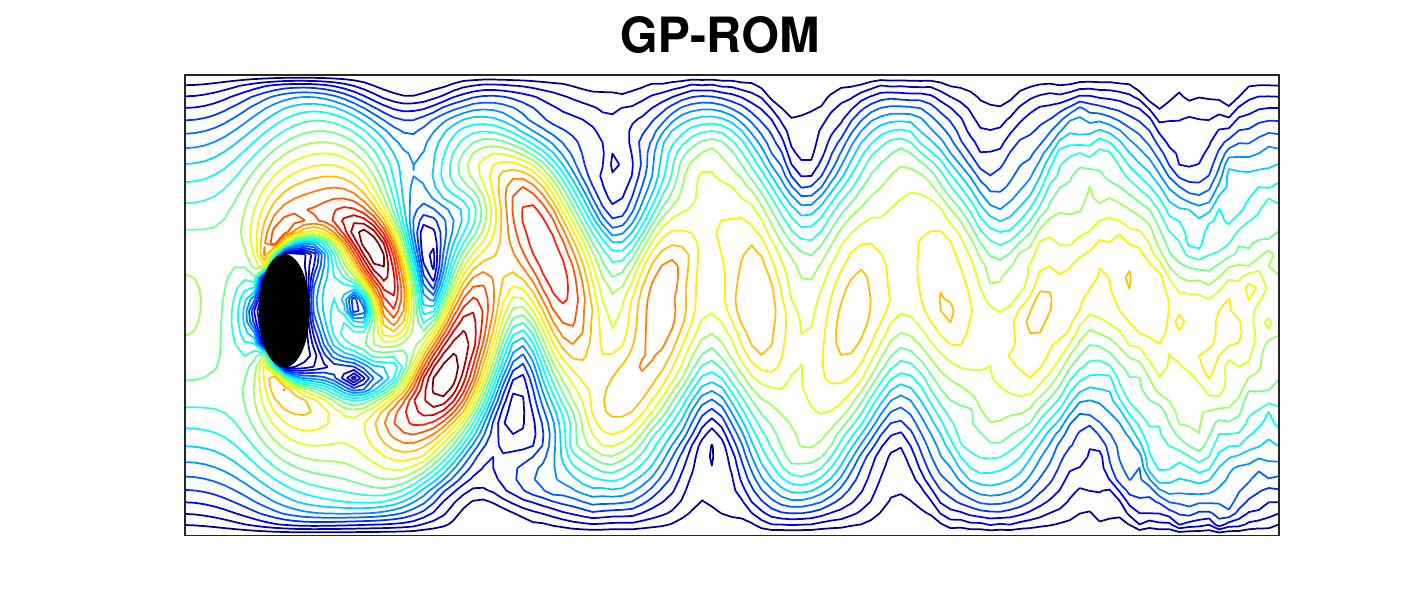}
\includegraphics[width=0.33\textwidth]{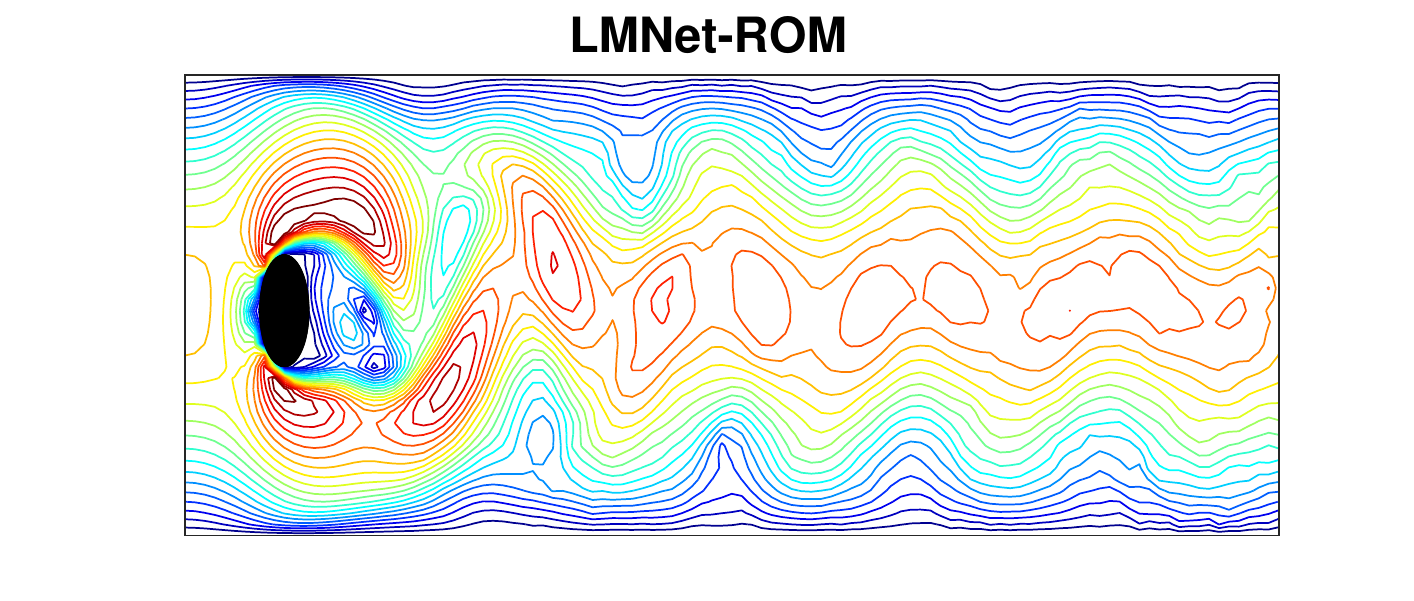}
\caption{ Vorticity prediction plots from the solution of LMNet-ROM (right) and GP-ROM (middle) with dimension $r=8$. The exact data (DNS) is plotted on the left. }
\label{fig:vorticity}
\end{figure*}

We also looked at the prediction of the time evolution of energy ($E(t_j) = \frac{1}{2}\|\bu^j\|_{L^2}$), vorticity construction, and drag coefficients. Fig.~\ref{fig:ener} shows the long-term prediction of energy travel and drag. The main observation is that the LMNet-ROM performs much better than GP-ROM. The energy and drag generated from GP-ROM have a huge deterioration when time evolves, which means the prediction is not accurate. Clearly, the prediction from LMNet-ROM is impressively good as it is stable and close to DNS, see Fig.~\ref{fig:ener}. Vorticity describes the local spinning motion of a fluid system.  Fig.~\ref{fig:vorticity} plots the vorticity construction from the velocity field around the cylinder at the end time $T=5$. As depicted in Fig.~\ref{fig:vorticity}, the LMNet-ROM correctly predicts the vortex street behind the cylinder while GP-ROM not. The above results are presented for dimension $r=8$, but similar results can be found for $r=4,6$. Overall, the long-term predictability of LMNet-ROM is much better than GP-ROM.  

\subsection{LMNet-ROM vs. Closure Models}
In this section, we present a preliminary numerical comparison between the LMNet-ROM and the closure models. Due to the wide class of stabilization closure models and the availability of their open source implementation, a thorough comparison between our new model and all other models is not practical. Therefore, we use the two most recent models (open source implemenation), Data-Driven-Filtered ROM (DDF-ROM) and Evolve-then-Filter ROM (EF-ROM), to illustrate the numerical comparison. The EF-ROM uses a two steps regularization strategy to improve the GP-ROM \cite{wells2017evolve}. The DDF-ROM solves an optimization problem to approximate the $\btau$ in system \eqref{eqn:rom-closure} \cite{xie2018data}. Table \ref{table:vs-closure} lists the average $L^2$ error from the three models with different dimensions. It surprisingly indicates that the closure models with mathematical methods cannot outperform the LMNet-ROM. This result is impressive, given that the LMNet-ROM is learned through the neural network from the data without acquiring any FOM information. 
 
  We note that the goal of both the stabilization closure models and our new model is to provide an accurate reduced dynamics to approximate the full system from different perspective. The former generally model the artificial term ($\btau$) in the system \eqref{eqn:rom-closure} physically or mathematically, whereas the latter use data to learn the dynamics \eqref{eqn:mnni-rom-online}. 
  We claim that the non-intrusive learning framework has the generality that can be applied to any nonlinear fluid system since it only requires the training data. 
  
\bgroup
\def\arraystretch{1.3}
\begin{table}[h!]
\caption{Average $L^2$ error from different models}
\vspace*{2mm}
\begin{tabular}{c|ccc}
\hline
\backslashbox{model}{dimension} & r=4 & r=6 & r=8  \\ \hline
         EF-ROM         & 1.23e-01 & 7.31e-02 & 1.84e-02 \\ \hline
         DDF-ROM        & 2.27e-01 & 1.14e-02 & 1.22e-02 \\ \hline
         \textbf{LMNet-ROM}  & \textbf{1.91e-03} & \textbf{3.57e-03} & \textbf{7.22e-03} \\ \hline
\end{tabular}
\label{table:vs-closure}
\end{table}
\egroup
\subsection{Computational Cost}
In this section, we discuss the computational efficiency of the proposed LMNet-ROM. The main computational cost of reduced order models is the offline computation since the cost of solving a small ODE system is negligible at the online stage. The FOM simulation (DNS) time is used as a benchmark to evaluate the performance of each model. The computation is carried out on a 64-bit Linux system with a single 2.70 GHz CPU. The DNS CPU time is 36828.53(s). Table \ref{table:cpu-time} lists the CPU time from each ROM and the associated speed-up factor. The LMNet-ROM time is only computed from one step AM method with one layer and 128 neurons, given the fact that this network architecture achieves good accuracy in the previous test. The result in Table \ref{table:cpu-time}reveals that the LMNet-ROM is more efficient compared to other models. This is a huge advantage of our new method since reducing the computational cost while maintaining good accuracy is the primary goal of reduced order modeling. 
\bgroup
\def\arraystretch{1.3}
\begin{table}[h!]
\centering
\caption{offline cost (second) and speed up factor from each ROM with dimension $r=8$}
\vspace*{2mm}
\begin{tabular}{c|cc}
\hline
         model             & cost & speed-up factor ($\frac{DNS}{ROM}$) \\ \hline
         GP-ROM             & 855.52s & 43.05\\ \hline
         EF-ROM         & 867.20s & 42.47\\ \hline
         DDF-ROM        & 6373.97s &  5.78\\ \hline
         \textbf{LMNet-ROM}  & \textbf{445.25(s)} & \textbf{82.71} \\ \hline
\end{tabular}
\label{table:cpu-time}
\end{table}
\egroup

\section{Conclusions and Outlook}
In this paper, we proposed a novel learning reduced order model framework for the numerical simulation of fluid flows. This framework was based on the recent development of linear multistep network architecture. We numerically studied the LMNet-ROM in the simulation of a 2D flow past a cylinder. The numerical results demonstrate that the LMNet-ROM was significantly more accurate than the GP-ROM in system approximation and long-term prediction. Furthermore, we compared the new model with the two most recent stabilization closure models, EF-ROM and DDF-ROM. The results show that our new method outperforms the two closure models. Overall, the LMNet-ROM beats the aforementioned models both in accuracy and computational efficiency, which provides a promising and encouraging approach in model reduction of fluid dynamics. However, the LMNet-ROM's potential still needs to be explored. We outline some research directions that could be pursued. 

Probably the most important next step is to study the parametrized system prediction of the LMNet-ROM. The current neural network is trained under the data-set given by one parameter value ($Re$) from the full order model of NSE. How does the LMNet-ROM predict systems with different parameter values, such as initial conditions and boundary conditions? Parametrized system prediction is a challenging problem in engineering applications. We hope to provide a systematic investigation with the new LMNet-ROM in later research.

Another important research direction is to improve the robustness of the model with respect to noise data. Table \ref{table:noise} shows the drawback of this model for high magnitude noisy ($\geq$1\%) data. We plan to address this issue by improving the neural network architecture. Also, regularization for preventing over-fitting needs to be fully studied in future research. 

Finally, the generality study of LMNet-ROM is worth investigation. Although we constructed and tested the LMNet-ROM in a fluid dynamics setting, the LMNet-ROM framework can be applied to any type of nonlinear partial differential equations (PDE) that is amenable to reduced order modeling. The only input needed in the LMNet-ROM framework is the data from the FOM of any system, see Algorithm~\ref{alg:nni-rom}. The LMNet-ROM procedure does not restrict it to the particular physical system modeled by the given nonlinear PDE. Since the LMNet-ROM is built by fully data-driven learning, we expect it to be successful in the numerical
simulation of general mathematical models (e.g., from elasticity or bioengineering).



\bibliography{xie}
\bibliographystyle{aaai}

\end{document}